\documentstyle[eqsecnum,aps,multicol,graphicx]{revtex}
\begin{document}
\draft
\title{Density-matrix functional theory of the Hubbard model:
An exact numerical study}
\author{R. L\'opez-Sandoval$^1$ and G. M. Pastor$^{1,2}$}
\address{$^1$Laboratoire de Physique Quantique, 
Unit\'e Mixte de Recherche 5626 du CNRS,\\
Universit\'e Paul Sabatier, 118 route de Narbonne, 
F-31062 Toulouse Cedex, France\\
$^2$Institut f\"ur Theoretische Physik, Freie Universit\"at Berlin, 
Arnimallee 14, D-14195 Berlin, Germany}

\date{\today}
\maketitle
        
\begin{abstract}  

A density functional theory for many-body lattice models is considered
in which the single-particle density matrix $\gamma_{ij}$ is the basic 
variable. Eigenvalue equations are derived for solving Levy's constrained 
search of the interaction  energy functional $W[\gamma_{ij}]$. 
$W[\gamma_{ij}]$ is expressed as the sum of 
Hartree-Fock energy $E_{\rm HF}[\gamma_{ij}]$ and the correlation
 energy $E_{\rm C}[\gamma_{ij}]$. 
Exact results are obtained for $E_{\rm C}(\gamma_{12})$ of 
the Hubbard model on various periodic lattices, where
$\gamma_{ij}= \gamma_{12}$ for all nearest neighbors $i$ and $j$. 
The functional dependence of $E_{\rm C}(\gamma_{12})$ is analyzed 
by varying the number of sites $N_a$, band filling $N_e$ and 
lattice structure. The infinite one-dimensional chain and
one-, two-, or three-dimensional finite clusters with periodic 
boundary conditions are considered. The properties of $E_{\rm C}(\gamma_{12})$ 
are discussed in the limits of weak ($\gamma_{12}\simeq \gamma_{12}^0$) and 
strong ($\gamma_{12}\simeq \gamma_{12}^\infty$) electronic correlations,
and in the crossover region 
($\gamma_{12}^\infty \le \gamma_{12} \le \gamma_{12}^0$).
Using an appropriate scaling we observe that 
$\varepsilon_{\rm C}(g_{12}) = E_{\rm C} / E_{\rm HF}$ has a
pseudo-universal behavior as a function of
$g_{12} = (\gamma_{12} - \gamma_{12}^{\infty}) / 
(\gamma_{12}^{0}-\gamma_{12}^{\infty})$. The fact that
 $\varepsilon_{\rm C}(g_{12})$ 
depends weakly on $N_a$, $N_e$ and lattice structure
suggests that the correlation energy of extended systems could be 
obtained quite accurately from finite cluster calculations.
Finally, the behavior of $E_{\rm C}(\gamma_{12})$ for repulsive ($U>0$) 
and attractive ($U<0$) interactions are contrasted. 

\end{abstract}
\vspace{0.5cm}

\pacs{Pacs numbers:\ 71.10.-w, 71.15.Mb, 71.10.Fd 
} 

\begin{multicols}{2}

\narrowtext

\section {Introduction}

Density functional theory (DFT) has been the subject of remarkable  
developments since its original formulation by Hohenberg and Kohn 
(HK) \cite{hk,ks}. After formal improvements, extensions, and an uncountable 
number of applications to a wide variety of physical problems, 
this theoretical approach has become the most efficient, 
albeit not infallible, method of determining the electronic properties 
of matter from first principles \cite{parr-book,gross-book}. The most 
important innovation of DFT, which is actually at the origin of its 
breakthrough, is to replace the wave function by the electronic 
density $\rho(\vec r)$ as the fundamental variable of the many-body 
problem. In practice, density functional (DF) calculations are 
largely based on the Kohn-Sham (KS) scheme that reduces the 
many-body $N$-particle problem to the solution of a set of self-consistent 
single-particle equations \cite{ks}. Although this transformation 
is formally exact, the implementations always require 
approximations, since the KS equations involve functional derivatives
of the unknown interaction energy $W[\rho(\vec{r})]$, usually expressed 
in terms of the exchange and correlation (XC) energy 
$E_{\rm XC}[\rho(\vec r)]$. Therefore, understanding the 
functional dependence of $E_{\rm XC}[\rho(\vec r)]$ and improving its 
approximations are central to the development of DF methods.
The currently most widespread {\em Ans\"atze} for $E_{\rm XC}[\rho(\vec r)]$ 
---the local density approximation (LDA) \cite{ks} with 
spin polarized \cite{hedin} and gradient corrected 
extensions \cite{gc}--- were originally derived from exact 
results for the homogeneous electron gas.  
It is one of the purposes of this paper to investigate the properties
of the interaction-energy functional from an 
intrinsically inhomogeneous point of view, namely, by considering
exactly solvable many-body lattice models.
 
Despite the remarkable success of the local spin density approximation, 
present DFT fails systematically in accounting for phenomena where strong 
electron correlations play a central role, for example, 
in heavy-fermion materials or high-$T_c$ superconductors. These 
systems are usually described by simplifying the low-energy 
electron dynamics using parameterized lattice models such as 
Pariser-Parr-Pople,\cite{ppp} Hubbard,\cite{hub} or Anderson\cite{andmod} 
models and related Hamiltonians \cite{fulde}. 
Being in principle an exact theory, the limitations of the DF 
approach have to be ascribed to the approximations used for exchange 
and correlation and not to the underlying HKS formalism. 
It would be therefore very interesting to extend the range of 
applicability of DFT to strongly correlated systems and to 
characterize the properties $E_{\rm XC}$ in the limit of strong 
correlations. Studies of the XC functional on simple models 
should provide useful insights for future extensions to realistic 
Hamiltonians. Moreover, taking into account the demonstrated power 
of the DF approach in {\em ab initio} calculations, one may also 
expect that a DFT with an appropriate $E_{\rm XC}$ could become an 
efficient tool for studying many-body models, a subject of theoretical 
interest on its own.

Several properties of DFT on lattice models have been already 
studied in previous works\cite{gunn86,god95,gunn95}. 
Gunnarsson and Sch\"onhammer were, to our knowledge, 
the first to propose a DF
approach on a semiconductor model in order to study the band-gap 
problem\cite{gunn86}. In this case the local site occupancies 
were treated as the basic variables. Some years later Schindlmayr and 
Godby \cite{god95} provided a different formulation of DFT on a lattice 
by considering as basic variables both diagonal elements $\gamma_{ii}$ 
and off-diagonal elements $\gamma_{ij}$ of the single-particle density 
matrix (see also \cite{gilb,donnelly,val}). 
Sch\"onhammer {\em et al.} then derived a more general 
framework that unifies the two previous approaches \cite{gunn95}.  
Using Levy's constrained search method \cite{levy} they showed 
that different basic variables and different $W$ functionals 
can be considered depending on the type of model or perturbation 
under study. Site occupations alone may be used as basic variables, 
if only the orbital energies are varied (i.e., if all hopping 
integrals $t_{ij}$ are kept constant for $i\not= j$). 
However, off-diagonal elements 
of the single-particle density matrix must be included explicitly if the 
functional $W$ is intended to be applied to more general situations 
involving different values of $t_{ij}$, for example, the Hubbard 
model on various lattice structures or for different interaction 
regimes, i.e., different $U/t$.  

In this paper we investigate the properties of Levy's interaction-energy 
functional $W$ as a function of $\gamma_{ij}$
by solving the constrained search minimization problem exactly.
In Sec.~\ref{sec:teo} the basic formalism of density-matrix
functional theory (DMFT) on  lattice 
models is recalled and the equations for determining 
$W[\gamma_{ij}]$ are derived.
Sec.~\ref{sec:res} presents and discusses exact results   
for the correlation energy $E_{\rm C}$ of the Hubbard model, which is
given by the difference between $W$ and the Hartree-Fock energy
 $E_{\rm HF}$. These are
obtained, either numerically for finite clusters with different 
lattice structures, or from the Bethe-Ansatz solution for the 
one-dimensional chain. Finally, Sec.~\ref{sec:conc} summarizes 
our conclusions and points out some relevant extensions.

\section{Theory}
\label{sec:teo}

In Sec.~\ref{sec:teolat} the main results of Levy's formulation 
of DMFT are presented in a form that is
appropriate for the study of model Hamiltonians such as the 
Hubbard model. Here, the hopping integrals $t_{ij}$ between sites 
(or orbitals) $i$ and $j$ play the role given in conventional DFT 
to the external potential $V_{ext}(\vec r)$. Consequently, 
the single-particle density matrix $\gamma_{ij}$ replaces the 
density $\rho(\vec r)$ as basic 
variable \cite{god95,gunn95,gilb,donnelly,val}. 
In Sec.~\ref{sec:teoex}, we derive equations that allow 
to determine Levy's interaction-energy functional 
$W[\gamma_{ij}]$ in terms 
of the ground-state energy of a many-body Hamiltonian 
with effective hopping integrals $\lambda_{ij}$ that depend 
implicitly on $\gamma_{ij}$.

\subsection{DMFT of lattice models}
\label{sec:teolat}

We consider the many-body Hamiltonian 
\begin{equation}
\label{eq:hamgen}
H = \sum_{ij\sigma}
t_{ij} \; \hat c_{i \sigma}^{\dagger} \hat c_{j \sigma} +
\frac{1}{2} \sum_{ijkl \atop{\sigma \sigma '}} 
V_{ijkl} \; \hat c_{i \sigma }^{\dagger} \hat c_{k   \sigma '}^{\dagger} 
            \hat c_{l \sigma '} \hat c_{j \sigma} \;,
\end{equation}
where  $\hat c_{i \sigma}^{\dagger}$  ($\hat c_{i \sigma}$) is the 
usual creation (annihilation) operator for an electron with spin 
$\sigma$ at site (or orbital) $i$. $H$ can be regarded as the second 
quantization of Schr\"odinger's equation on a basis \cite{foot_tij}.
However, in the present paper, the hopping integrals $t_{ij}$ 
and the interaction matrix elements $V_{ijkl}$ are taken as 
parameters to be varied independently. The matrix $t_{ij}$ defines 
the lattice (e.g., one dimensional chains, square or triangular 
two-dimensional lattices) and the range of single-particle interactions 
(e.g., up to first or second neighbors). From the {\em ab initio} perspective
$t_{ij}$ is given by the external potential and by the choice of the 
basis \cite{foot_tij}. $V_{ijkl}$ defines
the type of many-body interactions which may be repulsive (Coulomb like) 
or attractive (in order to simulate electronic pairing) and 
which are usually approximated as short ranged (e.g.,
intra-atomic). Eq.~(\ref{eq:hamgen}) is mainly used in this section 
to derive general results which can then be applied to 
various specific models by simplifying the interactions. 
A particularly relevant example, to be considered in some detail in 
Sec.~\ref{sec:res}, is the single-band Hubbard model with nearest 
neighbor (NN) hoppings \cite{hub}, which can be obtained from 
Eq.~(\ref{eq:hamgen}) by setting $t_{ij}= -t$ for $i$ and $j$ NN's, 
$t_{ij} = 0$ otherwise, and 
$V_{ijkl} = U \delta_{ij} \delta_{kl} \delta_{ik}$ \cite{ppp,fulde}. 

In order to apply DMFT to model Hamiltonians of the form 
(\ref{eq:hamgen}) we follow Levy's constrained search 
procedure \cite{levy} as proposed by Schindlmayr and 
Godby \cite{god95}. The ground-state energy is determined 
by minimizing the functional
\begin{equation}
\label{eq:efun}
E[\gamma_{ij}] = E_K[\gamma_{ij}] + W [\gamma_{ij}]
\end{equation}
with respect to the single-particle density matrix $\gamma_{ij}$. 
$E[\gamma_{ij}]$ is physically defined for all 
density matrices that can be written as
\begin{equation}
\label{eq:gamma} 
\gamma_{ij} = \sum_{\sigma}\gamma_{ij \sigma}=
\sum_\sigma \langle \Psi| c_{i \sigma }^{\dagger}c_{j \sigma }|\Psi \rangle 
\end{equation}
for all $i$ and $j$, where $|\Psi \rangle$ is an $N$-particle 
state. In other words, $\gamma_{ij}$ must derive from a physical state.
It is then said to be pure-state $N$-representable \cite{rep,foot_ens}.
The first term in Eq.~(\ref{eq:efun}) is given by
\begin{equation}
\label{eq:ek}
E_K = \sum_{ij} t_{ij}\gamma_{ij} \; .
\end{equation}
It includes all single-particle contributions and is usually 
regarded as the kinetic energy associated with the electronic 
motion in the lattice. Notice that Eq.~(\ref{eq:ek}) yields the 
exact kinetic energy for a given $\gamma_{ij}$. There are no 
corrections on $E_K$ to be included in other parts of the functional
as in the KS approach. The second term in Eq.~(\ref{eq:efun})
is the interaction-energy 
functional given by \cite{levy}
\begin{equation}
\label{eq:w}
W[ \gamma_{ij}] = 
min \left[\frac{1}{2} \sum_{ nmkl \atop {\sigma \sigma '} }  V_{nmkl} \;
\langle \Psi [ \gamma_{ij} ] | \;
\hat c_{n \sigma }^{\dagger} \hat c_{k \sigma '}^{\dagger}
\hat c_{l \sigma '}           \hat c_{m \sigma} 
\; |\Psi[\gamma_{ij}]\rangle  \right] \; .
\end{equation}
The minimization in Eq.~(\ref{eq:w}) implies a search over all 
$N$-particles states $| \Psi [\gamma_{ij}] \rangle$ 
that satisfy $\langle \Psi [\gamma_{ij}] | \; 
\sum_\sigma \hat c_{i \sigma }^{\dagger}\hat c_{j \sigma} 
\; |\Psi [\gamma_{ij}] \rangle = \gamma_{ij}$ for all $i$ and $j$.  
Therefore, $W[\gamma_{ij}]$ represents the minimum value of the 
interaction energy compatible with a given density matrix 
$\gamma_{ij}$. $W$ is usually expressed in terms of the Hartree-Fock energy 
\begin{equation}
\label{eq:HF}
E_{\rm HF}[\gamma_{ij}]=\frac{1}{2}\sum_{ijkl \atop{\sigma \sigma'}} 
V_{ijkl} \left(\gamma_{ij \sigma} \gamma_{kl \sigma'} 
- \delta_{\sigma\sigma '}  \gamma_{il \sigma} \gamma_{kj \sigma} 
\right) 
\end{equation}
\noindent and the correlation energy $E_{\rm C}[\gamma_{ij}]$ as
\begin{equation}
\label{eq:xc}
W[\gamma_{ij}] = E_{\rm HF}[\gamma_{ij}] + E_{\rm C}[\gamma_{ij}] \;.
\end{equation}                  
$W$ and $E_{\rm C}$ are universal functionals of $\gamma_{ij}$ in 
the sense that they are independent of $t_{ij}$, i.e., of the 
system under study. They depend on the considered interactions or 
model, as defined by $V_{ijkl}$, on the number of electrons $N_e$,
and on the structure of the many-body Hilbert space, as given by 
$N_e$ and the number of orbitals or sites $N_a$. Notice that $E_{\rm C}$ in
Eq.~(\ref{eq:xc}) does not include any exchange contributions. Given 
$\gamma_{ij}$ ($\gamma_{ij \sigma}=\gamma_{ij}/2$ in nonmagnetic cases) 
there is no need to approximate the exchange term, which is taken 
into account exactly by $E_{\rm HF}$ [Eq.~(\ref{eq:HF})]. 
Nevertheless, if useful in 
practice, it is of course  possible to split $W$ in the Hartree energy 
$E_{\rm H}$ and the exchange and correlation energy $E_{\rm XC}$ is a similar 
way as in the KS approach. 

The variational principle results from the following two 
relations \cite{levy}:
\begin{equation}
\label{eq:e0<}
E_{gs} \le \sum_{ij}t_{ij} \gamma_{ij} + W[\gamma_{ij}]  
\end{equation}
for all pure-state $N$-representable $\gamma_{ij}$ \cite{rep}, and
\begin{equation}
\label{eq:e0=}
E_{gs} = \sum_{ij} t_{ij} \gamma_{ij}^{gs} + W[\gamma_{ij}^{gs}] \;,
\end{equation}
where  $E_{gs} = \langle \Psi_{gs} | H |\Psi_{gs} \rangle$ refers to the 
ground-state energy and $\gamma_{ij}^{gs} = \langle \Psi_{gs} | 
\sum_\sigma \hat c_{i \sigma }^\dagger \hat c_{j \sigma } |\Psi_{gs} \rangle$ 
to the ground-state single-particle density matrix.

As already pointed out in previous works \cite{god95,gunn95}, 
$W$ and $E_{\rm C}$ depend in general on both diagonal elements $\gamma_{ii}$
and off-diagonal elements $\gamma_{ij}$ of the density-matrix,
since the hopping integrals $t_{ij}$ are non local in the 
sites. The situation is similar to the DF approach proposed by 
Gilbert for the study of non-local potentials 
$V_{ext}(\vec r, {\vec r}\;\!')$ as those appearing in the theory of 
pseudo-potentials \cite{gilb,donnelly,val}. 
A formulation of DFT on a lattice only in terms of $\gamma_{ii}$ 
would be possible if one would restrict oneself to a family of 
models with constant $t_{ij}$ for $i\not= j$. However, in this case 
the functional $W[\gamma_{ii}]$ would depend on the actual value of 
$t_{ij}$ for $i\not= j$ \cite{gunn95}.

The functional $W[\gamma_{ij}]$, valid for all lattice 
structures and for all types of hybridizations, 
can be simplified at the expense of universality 
if the hopping integrals are short ranged. For example, if only
NN hoppings are considered, the kinetic energy $E_K$ is independent
of the density-matrix elements between sites that are not NN's.
Therefore, the constrained search in Eq.~(\ref{eq:w}) may restricted to the 
$| \Psi [\gamma_{ij}] \rangle$ that satisfy $\langle \Psi [\gamma_{ij}] |\; 
\sum_\sigma \hat c_{i \sigma }^{\dagger}\hat c_{j \sigma} 
\;|\Psi [\gamma_{ij}] \rangle = \gamma_{ij}$ 
only for $i=j$ and for NN $ij$. In this way the number 
of variables in $W[\gamma_{ij}]$ is reduced significantly
rendering the interpretation of the functional dependence 
far simpler. While this is a great practical 
advantage, it also implies that $W$ and $E_{\rm C}$ lose their 
universal character since the dependence on the NN $\gamma_{ij}$
is now different for different lattices. In Sec.~\ref{sec:res} results 
for one-, two-, and three-dimensional lattices with NN hoppings are 
compared in order to quantify this effect.
 
For the applications in Sec.~\ref{sec:res} we shall consider 
the single-band Hubbard model with NN hoppings, which in the 
usual notation is given by\cite{hub} 
\begin{equation}
\label{eq:hamhub}
H = -t  \sum_{\langle i,j\rangle \sigma} 
\hat c^{\dagger}_{i \sigma} \hat c_{j \sigma} +
U \sum_i  \hat n_{i \downarrow} \hat n_{i\uparrow} \; .
\end{equation}
In this case the interaction energy functional reads
\begin{equation}
\label{eq:whub}
W[\gamma_{ij}] = 
min \left[ U \sum_l  \langle \Psi [\gamma_{ij}] |\; 
\hat n_{l\uparrow} \hat n_{l\downarrow} 
\;|\Psi [\gamma_{ij}] \rangle  \right] \; ,
\end{equation}
where the minimization is performed with respect to all $N$-particle 
$|\Psi [\gamma_{ij}]\rangle$ satisfying $\langle \Psi [\gamma_{ij}] | 
\sum_\sigma \hat c_{i \sigma }^{\dagger}\hat c_{j \sigma} |
\Psi [\gamma_{ij}] \rangle = \gamma_{ij}$
for $i$ and $j$ NN's. If the interactions are repulsive ($U>0$)
$W[\gamma_{ij}]$ represents the minimum average number of 
double occupations which can be obtained for a given degree 
of electron delocalization, i.e., for a given value of $\gamma_{ij}$.
For attractive interactions ($U<0$) double occupations are 
favored and $W[\gamma_{ij}]$ corresponds to the maximum of
$ \sum_l  \langle \hat n_{l\uparrow} \hat n_{l\downarrow}\rangle$
for a given $\gamma_{ij}$.

\subsection{Exact XC energy functional}
\label{sec:teoex}

In order to determine $E_{\rm C}[\gamma_{ij}]$ and $W[\gamma_{ij}]$ 
we look for the extremes of
\begin{eqnarray}
\label{eq:f}
F &=& \sum_{ ijkl \atop {\sigma \sigma '} }  \left [ V_{ijkl}  
\langle\Psi| \hat c_{i \sigma }^{\dagger} \hat c_{k \sigma '}^{\dagger}
\hat c_{l \sigma } \hat c_{j \sigma} |\Psi\rangle \right]
 \; + \;\varepsilon \; \Big(1 - \langle \Psi |\Psi \rangle \Big) 
 \;  \nonumber  \\
 &+& \sum_{i,j} \lambda_{ij} \;
 \Big( \langle\Psi| 
       \sum_\sigma \hat c^\dagger_{i\sigma} \hat c_{j \sigma} |
       \Psi\rangle  -  \gamma_{ij} \Big) 
\end{eqnarray}
with respect to $|\Psi\rangle $. Lagrange multipliers $\varepsilon$ 
and $\lambda_{ij}$ have been introduced to enforce the normalization of 
$|\Psi\rangle$ and the conditions on the representability of 
$\gamma_{ij}$. Derivation with respect to $ \langle\Psi|$, $\varepsilon$
and $\lambda_{ij}$ yields the eigenvalue equations 
\begin{equation}
\label{eq:evgen}
\sum_{ij\sigma} \lambda_{ij} \; \hat c^{\dagger}_{i \sigma} \hat c_{j \sigma} 
\; |\Psi \rangle  + 
\sum_{ ijkl \atop {\sigma \sigma '} } V_{ijkl}\;  
\hat c_{i \sigma }^{\dagger} \hat c_{k \sigma '}^{\dagger}
\hat c_{l \sigma } \hat c_{j \sigma} \; |\Psi\rangle  
= \varepsilon \; | \Psi \rangle \; , 
\end{equation}
and the auxiliary conditions $\langle \Psi | \Psi \rangle = 1$ and 
 $\gamma_{ij} = \langle\Psi| 
\sum_\sigma \hat c^\dagger_{i\sigma} \hat c_{j \sigma}|\Psi\rangle$. 
The Lagrange multipliers $\lambda_{ij}$ play the role of hopping
integrals to be chosen in order that $|\Psi\rangle$ yields the given 
$\gamma_{ij}$. The pure-state representability of $\gamma_{ij}$
ensures that there is always a solution \cite{rep}. In practice, however,
one usually varies $\lambda_{ij}$ in order to scan the domain of 
representability of $\gamma_{ij}$. 
For given $\lambda_{ij}$, the eigenstate $|\Psi_0 \rangle$
corresponding to the lowest eigenvalue of Eq.~(\ref{eq:evgen}) 
yields the minimum $W[\gamma_{ij}]$ for $\gamma_{ij} = \langle\Psi_0| 
\sum_\sigma \hat c^\dagger_{i\sigma} \hat c_{j \sigma}|\Psi_0\rangle$. 
Any other $|\Psi\rangle$ satisfying $\gamma_{ij} = \langle\Psi| 
\sum_\sigma \hat c^\dagger_{i\sigma} \hat c_{j \sigma}|\Psi\rangle$
would have higher $\varepsilon$ and thus higher $W$. 
The subset of $\gamma_{ij}$ which are representable by a ground-state
of Eq.~(\ref{eq:evgen}) is the physically relevant one, since it 
necessarily includes the absolute minimum $\gamma_{ij}^{gs}$ of 
$E[\gamma_{ij}]$. Nevertheless, it should be noted that 
pure-state representable $\gamma_{ij}$ may be considered that can only 
be represented by excited states or by linear combinations of
eigenstates of Eq.~(\ref{eq:evgen}). In the later case, 
$\lambda_{ij} = 0 \; \forall ij$, and $|\Psi_0\rangle$ is an 
eigenstate of the interaction term with lowest eigenvalue.  
Examples shall be discussed in Sec.~\ref{sec:res}.

For the Hubbard model Eq.~(\ref{eq:evgen}) reduces to 
\begin{equation}
\label{eq:evhub}
\sum_{\langle ij\rangle \atop \sigma} \lambda_{ij} \;
\hat c^{\dagger}_{i\sigma}\hat c_{j\sigma} \; |\Psi\rangle \; + \;
U \sum_i \hat n_{i\uparrow} \hat n_{i\downarrow} \; |\Psi\rangle  
= \varepsilon  \; |\Psi\rangle \; .
\end{equation}
This eigenvalue problem can be solved numerically for clusters
with different lattice structures and periodic boundary conditions.
In this case we expand $| \Psi [\gamma_{ij}] \rangle$
in a complete set of basis states $\vert\Phi_m\rangle$ which have 
definite occupation numbers $\nu^m_{i \sigma}$ at all orbitals $i\sigma$ 
($\hat n_{i\sigma} \vert\Phi_m\rangle = \nu_{i\sigma}^m 
\vert\Phi_m\rangle$ 
with $\nu_{i\sigma}^m = 0$ or $1$). The values of $\nu^m_{i \sigma}$ 
satisfy the usual conservation of the number of electrons 
$N_e  = N_{e\uparrow} + N_{e\downarrow}$ and of the $z$ component 
of the total spin $S_z = (N_{e\uparrow} - N_{e\downarrow})/2$, where
$N_{e\sigma} = \sum_{i} \nu^m_{i \sigma}$. For not too large clusters, 
the lowest energy $|\Psi_0 [\gamma_{ij}] \rangle$ 
---the ground state of Eq.~(\ref{eq:evhub})--- 
can be determined by sparse-matrix diagonalization procedures,
for example, by using Lanczos iterative method \cite{lan}.
$| \Psi_0 [\gamma_{ij}] \rangle$ is calculated in the subspace 
of minimal $S_z$ since this ensures that there are no {\em a priori\/} 
restrictions on the total spin $S$. In addition, spin-projector operators 
may be used to study the dependence of $E_{\rm C}(\gamma_{12})$ on $S$.

For a one-dimensional (1D) chain with NN hoppings $t_{ij} = t$, 
translational symmetry implies equal density-matrix elements 
$\gamma_{ij}$ between NN's. Therefore, one may set 
$\lambda_{ij} = \lambda$ for all NN $ij$, and then 
Eq.~(\ref{eq:evhub}) has the same form as the 1D Hubbard 
model for which Lieb and Wu's exact solution is 
available \cite{lieb-wu}. 
In this case the lowest eigenvalue $\varepsilon$ is determined following 
the work by Shiba \cite{shiba}. The coupled Bethe-Ansatz equations
are solved as a function of $\lambda$, band-filling $n=N_e/N_a$, and 
for positive and negative $U$, by means of a simple iterative procedure.

\section{Results and Discussion}
\label{sec:res}

In this section we present and discuss exact results for the 
correlation energy functional $E_{\rm C}[\gamma_{ij}]$ of 
the single-band Hubbard Hamiltonian with nearest neighbor 
hoppings \cite{hub}. Given the lattice structure, $N_a$ and $N_e$,
the model is characterized by the 
dimensionless parameter $U/t$ which measures the competition between 
kinetic and interaction energies [see Eq.~(\ref{eq:hamhub})].
$U>0$ corresponds to the usual intra-atomic repulsive Coulomb 
interaction, while the attractive case ($U<0$) simulates 
intra-atomic pairing of electrons. 

\subsection{Repulsive interaction $U>0$}

\begin{figure}[x]
\centerline{\resizebox{8.2cm}{10.2cm}{\includegraphics{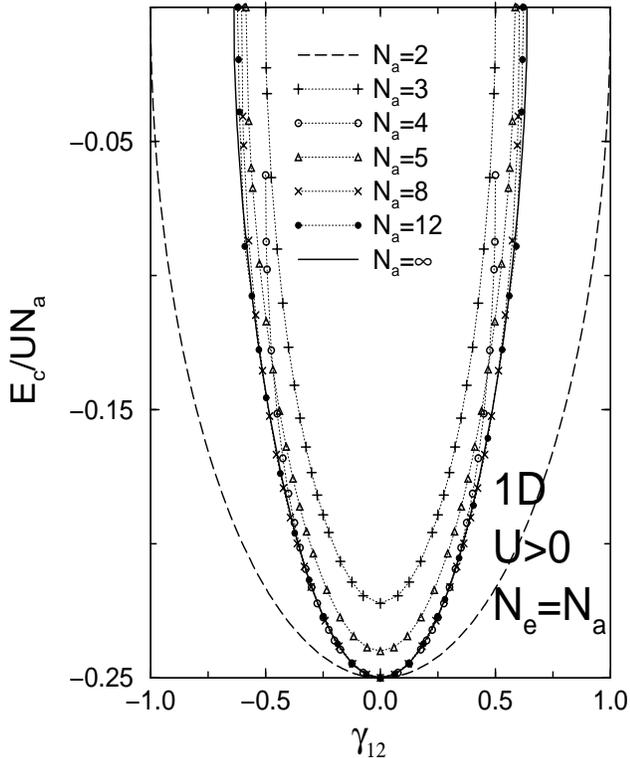}}}
\caption{
Correlation energy $E_{\rm C}$ of the 
Hubbard model on one-dimensional rings with $N_a$ sites
and $N_e = N_a$ electrons as a function of the density-matrix 
element or bond order $\gamma_{12}$ between nearest neighbors (NN).
$\gamma_{ij} = \gamma_{12}$ for all NN $ij$. $U$ refers to the 
intra-atomic Coulomb repulsion [$U>0$, see Eq.~(\protect\ref{eq:hamhub})]
\protect\cite{foot_e-hole}. 
}
\label{fig:xcanft}
\end{figure}

In Fig.~\ref{fig:xcanft} the correlation energy $E_{\rm C}$ of the 
one-dimensional (1D) Hubbard model is shown for half-band filling 
($N_e = N_a$) as a function of the density-matrix element or 
bond order $\gamma_{12}$ between NN's. $\gamma_{ij} = \gamma_{12}$ 
for all NN's $i$ and $j$. Results are given for rings of finite 
length $N_a$ as well as for the infinite chain.
Several general qualitative features may be identified.
First of all we observe that on bipartite lattices \cite{foot_bip}
$E_{\rm C}(\gamma_{12}) = E_{\rm C}(-\gamma_{12})$, since the sign 
of the NN bond order can be changed without affecting the interaction 
energy $W(\gamma_{12})$ by changing the phase of the local orbitals 
at one of the sublattices ($c_{i\sigma} \to -c_{i\sigma}$ 
for $i\in A$ and $c_{j\sigma}$ unchanged for $j\in B$, 
where $A$ and $B$ refer to the sublattices).
Let us recall that the domain of definition of $E_{\rm C}(\gamma_{12})$
is limited by the pure-state representability of $\gamma_{ij}$.
The upper bound $\gamma_{12}^{0+}$ and the lower bound 
$\gamma_{12}^{0-}$ for $\gamma_{12}$ 
($\gamma_{12}^{0+}= - \gamma_{12}^{0-} =  \gamma_{12}^0$ on 
bipartite lattices) are the extreme values of the bond order 
between NN's on a given lattice and for given $N_a$ and $N_e$ 
($\gamma_{ij} = \gamma_{12}$ for all NN $ij$). They represent the 
maximum degree of electron delocalization. $\gamma_{12}^{0+}$ and 
$\gamma_{12}^{0-}$ correspond to the extremes of the kinetic 
energy $E_K$ [$E_K = \sum_{\langle ij\rangle} t _{ij} \gamma_{ij} 
=  (z N_a/2) t \gamma_{12}$, where $z$ is the coordination number]
and thus to the ground state of the Hubbard model for $U=0$ 
[$\gamma_{12}^{0+}$ for $t>0$ and $\gamma_{12}^{0-}$ for $t<0$,
see Eq.~(\ref{eq:hamhub})]. For $\gamma_{12} = \gamma_{12}^{0}$ 
the underlying electronic state $|\Psi_0\rangle$ is usually a single 
Slater determinant and therefore $E_{\rm C}(\gamma_{12}^0)=0$. 
In other words, the correlation energy vanishes 
as expected in the fully delocalized limit \cite{foot_U=0}.
As $|\gamma_{12}|$ decreases 
$E_{\rm C}$ decreases ($E_{\rm C} < 0$) since correlations 
can reduce the Coulomb energy more and more efficiently as the 
electrons localize. $E_{\rm C}$ is minimum in the strongly correlated 
limit $\gamma_{12} = \gamma_{12}^\infty$. For half-band filling 
this corresponds to a fully localized electronic state 
($\gamma_{12}^\infty =0$). Here, $E_{\rm C}$ cancels out the 
Hartree-Fock energy $E_{\rm HF}$ and the Coulomb energy $W$ vanishes 
($E_{\rm C}^\infty = -E_{\rm HF}$) \cite{foot_EH}.
The ground-state values of $\gamma_{12}^{gs}$ and $E_{gs}$ for a
given $U/t$ result from the competition between lowering $E_{\rm C}$ 
by decreasing $\gamma_{12}$ and lowering $E_K$ by increasing it ($t>0$). 
The divergence of $\partial E_{\rm C} / \partial\gamma_{12}$ for
$\gamma_{12} = \gamma_{12}^0$ is a necessary 
condition in order that $\gamma_{12}^{gs} < \gamma_{12}^0$ for 
arbitrary small $U>0$. On the other side, for small $\gamma_{12}$,
we observe that $(E_{\rm C}+E_{\rm HF}) \propto \gamma_{12}^2$. This implies
that for $U/t\gg 1$, $\gamma_{12}^{gs} \propto t/U$ and 
$E_{gs}\propto t^2/U$, a well known result in the Heisenberg 
limit of the Hubbard model ($N_e=N_a$) \cite{fulde}.

\begin{figure}[x]
\centerline{\resizebox{8.2cm}{10.2cm}{\includegraphics{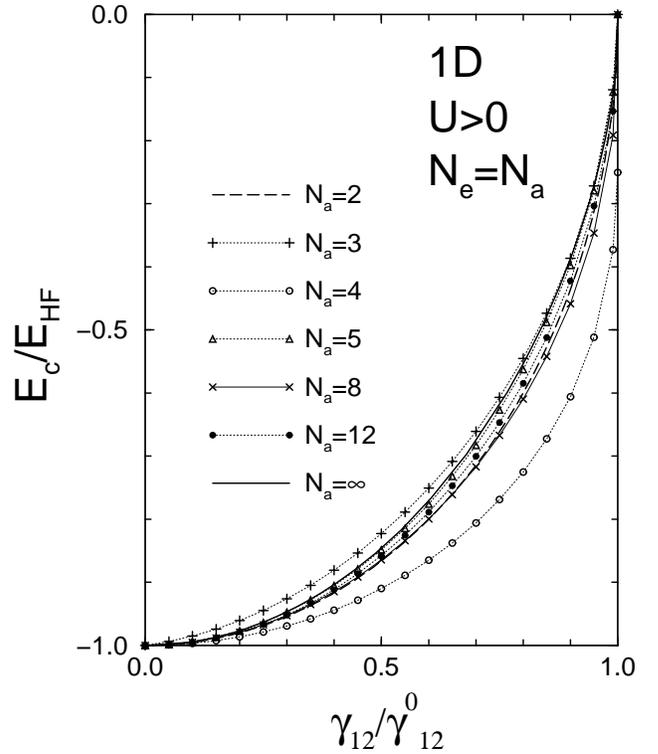}}}
\caption{
Correlation energy $E_{\rm C}$ in units of the Hartree-Fock
energy $E_{\rm HF}$ \protect\cite{foot_EH} for the Hubbard model on 
one-dimensional rings. 
Results are given as a function of $\gamma_{12} / \gamma_{12}^0$,
where $\gamma_{12}^0$ stands for the NN ground-state bond order 
in the uncorrelated limit ($U=0$). $\gamma_{ij} = \gamma_{12}$ for all 
NN $ij$. $N_a$ refers to the number of sites, and $N_e=N_a$ to the number 
of electrons. $E_{\rm C}(\gamma_{12}) = E_{\rm C}(-\gamma_{12})$, 
see Fig.~\protect\ref{fig:xcanft}.
}
\label{fig:nxcanft}
\end{figure}

A more quantitative analysis of $E_{\rm C}(\gamma_{12})$ and in 
particular the comparison of results for different $N_a$ 
is complicated by the size dependence of $\gamma_{12}^0$ and $E_{\rm HF}$.
It is therefore useful to measure $E_{\rm C}$ in units of the Hartree-Fock 
energy and to bring the domains of representability to a common range
by considering $\varepsilon_{\rm C} = E_{\rm C} / E_{\rm HF}$ as 
a function of $g_{12}=\gamma_{12}/ \gamma_{12}^0$. 
Fig.~\ref{fig:nxcanft} shows that $\varepsilon_{\rm C}(g_{12})$ 
has approximately the same behavior for all considered $N_a$. 
Finite size effects are small except for the very small sizes. 
The largest deviations from the common trend are found for 
$N_a = N_e = 4$. Here we observe a discontinuous drop of   
$\varepsilon_{\rm C}$ for  $g_{12}=1$ ($g_{12}< 1$)
which is due to the degeneracy of the single-particle 
spectrum. In fact in this case two of the four electrons occupy a doubly 
degenerate state in the uncorrelated limit and the minimum interaction  
energy $W(\gamma_{12})$ does not correspond to a single-Slater-determinant 
state even for $\gamma_{12}=\gamma_{12}^0$ \cite{foot_W0}. 
As $N_a$ increases 
$\varepsilon_{\rm C}(g_{12})$ approaches the infinite-length limit with 
alternations around the $N_a=\infty$ curve. 
The strong similarity between $\varepsilon_{\rm C}(g_{12})$ for small $N_a$
and for $N_a=\infty$ is a remarkable result. It suggests that 
good approximations for $E_{\rm C}(\gamma_{12})$ in extended systems could be 
derived from finite cluster calculations.

\begin{figure}[x]
\centerline{\resizebox{8.2cm}{10.2cm}{\includegraphics{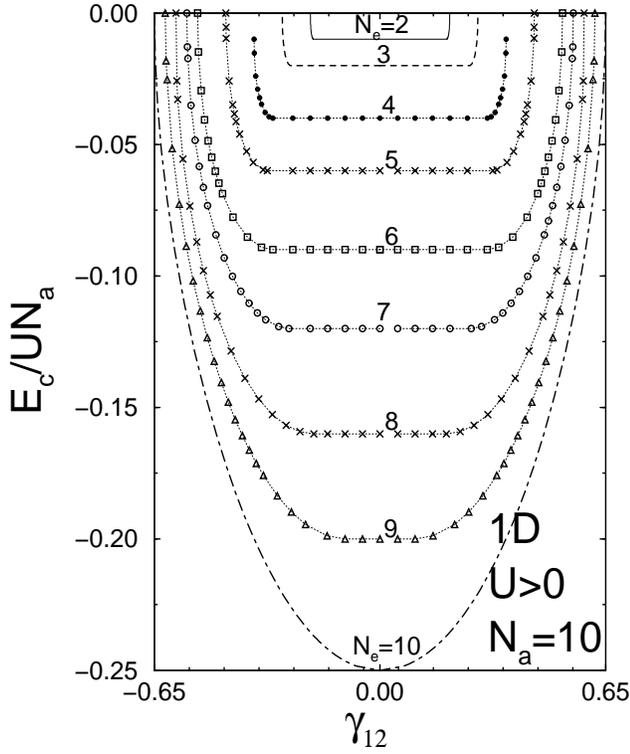}}}
\caption{
Band-filling dependence of the correlation energy 
$E_{\rm C}(\gamma_{12})$ of the one-dimensional Hubbard model for 
$N_a= 10$ sites. $\gamma_{ij} = \gamma_{12}$ for all NN $ij$. 
$N_e$ refers to the number of electrons and $U$ to the intra-atomic 
Coulomb repulsion. On bipartite lattices
$E_{\rm C}(\gamma_{12},N_e) = E_{\rm C}(-\gamma_{12}, N_e) = 
E_{\rm C}(\gamma_{12}, 2 N_a - N_e)$ \protect\cite{foot_e-hole}. 
}
\label{fig:xcanfl1}
\end{figure}

Fig.~\ref{fig:xcanfl1} shows the band-filling dependence of 
$E_{\rm C}(\gamma_{12})$ in a 10-site 1D Hubbard ring. 
Results are given for $N_e \le N_a$,  since for $N_e \ge N_a$,
$E_{\rm C}(\gamma_{ij}, N_e) = E_{\rm C}(-\gamma_{ij}, 2 N_a - N_e)$ 
as a result of electron-hole symmetry \cite{foot_e-hole}. 
Although $E_{\rm C}(\gamma_{12})$ depends strongly on $N_e$, 
several qualitative properties are shared by all band fillings:
(i) As in the half-filled band case, the domain of representability 
of $\gamma_{12}$ is bound by the bond orders in the uncorrelated limits. 
In fact, $\gamma_{12}^{0-}\le \gamma_{12} \le \gamma_{12}^{0+}$, where  
$\gamma_{12}^{0+}$ ($\gamma_{12}^{0-}$) corresponds to the ground state 
of the $U=0$ tight-binding model for $t>0$ ($t<0$).
On bipartite lattices $\gamma_{12}^{0+} = -\gamma_{12}^{0-} = \gamma_{12}^0$.
Notice that $\gamma_{12}^0$ increases monotonously with $N_e$ as the 
single-particle band is filled up. This is an important contribution to 
the band-filling dependence of $E_{\rm C}$ (see Fig.~\ref{fig:xcanfl1}).
(ii) In the delocalized limit, $E_{\rm C}(\gamma_{12}^0)=0$ 
for all the $N_e$ for which  $W(\gamma_{12}^0)$ derives from a 
single Slater determinant \cite{foot_U=0}. Moreover, the divergence of 
$\partial E_{\rm C} / \partial\gamma_{12}$ for 
$\gamma_{12}= \gamma_{12}^0$ indicates that 
$\gamma_{12}^{gs} < \gamma_{12}^0$ for arbitrary small $U>0$, as 
expected from perturbation theory.
(iii) Starting from $\gamma_{12} = \gamma_{12}^0$, $E_{\rm C}(\gamma_{12})$ 
decreases with decreasing $\gamma_{12}$ reaching its lowest possible 
value $E_{\rm C}^\infty = -E_{\rm HF}$ for 
$\gamma_{12} = \gamma_{12}^{\infty +}$ ($N_e \le N_a$). The same behavior 
is of course observed for $\gamma_{12}<0$. In particular, 
$E_{\rm C} = -E_{\rm HF}$ also for $\gamma_{12}=\gamma_{12}^{\infty -}$. 
As shown in Fig.~\ref{fig:xcanfl1}, $E_{\rm C}^\infty$ decreases rapidly 
with increasing $N_e$, since $E_{\rm HF}$ increases quadratically with the 
electron density \cite{foot_EH}. 
(iv) On bipartite lattices $\gamma_{12}^{\infty +}=-\gamma_{12}^{\infty -} =
\gamma_{12}^{\infty}$, while on non-bipartite structures 
one generally has $|\gamma_{12}^{\infty +}| \ne |\gamma_{12}^{\infty -}|$, 
since the single-particle spectrum is different for positive
and negative energies. The decrease of $E_{\rm C}$ with decreasing   
$|\gamma_{12}|$ shows that the reduction of the Coulomb energy due 
to correlations is done at the expense of kinetic energy or 
electron delocalization, as already discussed for 
$N_e = N_a$ (Fig.~\ref{fig:xcanft}). 
(v) $\gamma_{12}^{\infty}>0$ for all $N_e < N_a$    
($\gamma_{12}^{\infty}=0$ for $N_e = N_a$). $\gamma_{12}^{\infty}$ 
represents the largest NN bond order that can be constructed 
under the constraint of vanishing Coulomb repulsion energy. 
A lower bound for $\gamma_{12}^{\infty}$ is given by the bond 
order $\gamma_{12}^{FM}$ in the fully-polarized ferromagnetic state 
($\gamma_{12}^{\infty} \ge \gamma_{12}^{FM}$). This is obtained 
by occupying the lowest single-particle states with all electrons of the 
same spin ($N_e \le N_a$). Therefore, $\gamma_{12}^{FM}$  
increases with $N_e$ for $N_e \le N_a/2$ and then decreases 
for $N_a/2<N_e \le N_a$ reaching  $\gamma_{12}^{FM}=0$ at half-band 
filling ($\gamma_{12}^{FM}>0$ for $N_e<N_a$). In this way the
non-monotonous dependence of $\gamma_{12}^{\infty}$ on $N_e$ 
can be explained (see Fig.~\ref{fig:xcanfl1}). 
(vi) The correlation energy is 
constant and equal to $-E_{\rm HF}$ for $\gamma_{12}^{\infty -}\le\gamma_{12}
\le \gamma_{12}^{\infty +}$. These values of $\gamma_{12}$ can never 
correspond to the ground-state energy of the Hubbard model, since 
in this range increasing $\gamma_{12}$ always lowers the kinetic 
energy ($t>0$) without increasing the Coulomb repulsion
($\gamma_{12}^{\infty } \le \gamma_{12}^{gs} \le \gamma_{12}^{0}$). 
For $\gamma_{12}^{\infty -} < \gamma_{12} < \gamma_{12}^{\infty +}$, 
$\gamma_{12}$ cannot be represented by a ground state of 
Eq.~(\ref{eq:evhub}). In this range $\gamma_{12}$ can be derived from a 
linear combination of states having minimal Coulomb repulsion \cite{foot_rep}.

\begin{figure}[x]
\centerline{\resizebox{8.2cm}{10.2cm}{\includegraphics{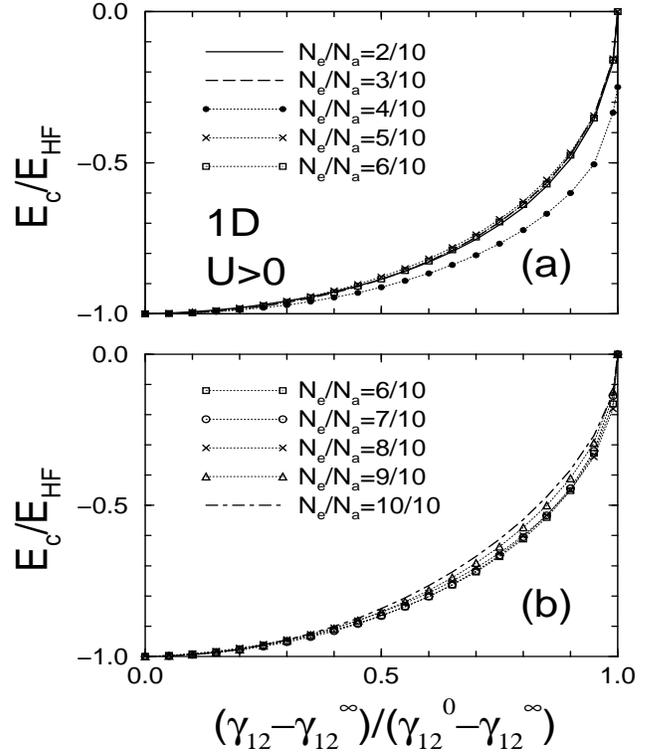}}} 
\caption{
Correlation energy $E_{\rm C}$ in units of the Hartree-Fock
energy $E_{\rm HF}$ \protect\cite{foot_EH} for the one-dimensional 
Hubbard model on a 1D 10-site ring. Results are given as a function 
of the degree of delocalization 
$g_{12}=(\gamma_{12} - \gamma_{12}^\infty) / 
(\gamma_{12}^0 - \gamma_{12}^\infty)$, 
where $\gamma_{12}^0$ refers to the NN bond order in the uncorrelated
ground-state ($U=0$) and $\gamma_{12}^\infty$ to the NN bond order 
in the strongly correlated limit ($U/t \to \infty$). 
As in Fig.~\protect\ref{fig:xcanfl1}, different band-fillings 
$N_e/N_a$ are considered: (a) $N_e\le 6$ and (b) $N_e \ge 6$.
}
\label{fig:nxcanfl1}
\end{figure}

\begin{figure}[x]
\centerline{\resizebox{8.2cm}{10.2cm}{\includegraphics{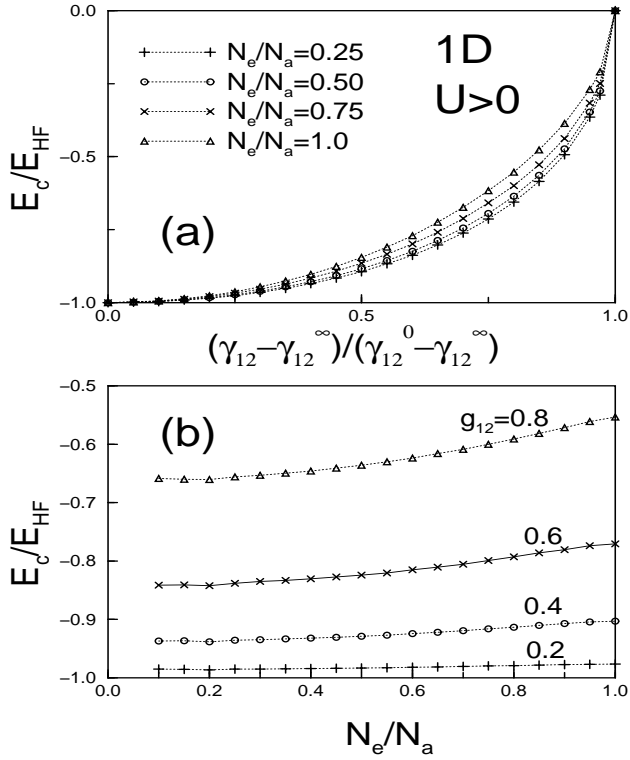}}}
\caption{
Correlation energy $E_{\rm C}$ of the Hubbard model on 
the infinite one-dimensional chain. Results are given for 
$\varepsilon_{\rm C}=E_{\rm C} / E_{\rm HF}$ 
($E_{\rm HF}=$ Hartree-Fock energy) as a function of (a) $g_{12}= 
(\gamma_{12} - \gamma_{12}^\infty) / (\gamma_{12}^0 - \gamma_{12}^\infty)$,
and (b) band filling $N_e/N_a$ 
(see caption of Fig.~\protect\ref{fig:nxcanfl1}).
}
\label{fig:1Dinf}
\end{figure}

In order to compare the functional dependences of the correlation energy 
for different band fillings, it is useful to scale $E_{\rm C}$ in 
units of the Hartree-Fock energy and to bring the relevant domains 
$\gamma_{12}^{\infty } \le \gamma_{12} \le \gamma_{12}^{0}$  of 
 different $N_e$ to a common range. In Fig.~\ref{fig:nxcanfl1},
 $\varepsilon_{\rm C}=E_{\rm C}/E_{\rm HF}$ is shown 
as a function of $g_{12} = (\gamma_{12}-\gamma_{12}^{\infty}) / 
(\gamma_{12}^{0}-\gamma_{12}^{\infty})$. We  observe that the 
results for $\varepsilon_{\rm C}(g_{12})$ are remarkably similar 
for all band-fillings. The largest deviations from the common 
trend are found for $N_e=4$. As already discussed for $N_a=N_e=4$, 
this anomalous behavior is  related to the degeneracy 
of the single-particle  spectrum and to the finite size of system. 
Fig.~\ref{fig:nxcanfl1} shows that for the Hubbard model
the largest part of the dependence of $E_{\rm C}(\gamma_{12})$ on 
band filling comes from $E_{\rm HF}$, $\gamma_{12}^{0}$ and 
$\gamma_{12}^{\infty}$. Similar conclusions are derived from 
the results for the infinite 1D chain presented in Fig.~\ref{fig:1Dinf}. 
For a given  $g_{12}$, $\varepsilon_{\rm C}(g_{12})$ depends weakly 
on $N_e/N_a$ if the carrier density is low ($N_e/N_a \le 0.4$), 
and tends to increase 
as we approach half-band filling [see Fig.~\ref{fig:1Dinf}(b)]. For high 
carrier densities it become comparatively more difficult 
to minimize the Coulomb energy for a given degree of delocalization 
$g_{12}$. The effect is most  pronounced for $g_{12}\simeq 0.8$--$0.9$, 
i.e., close to the uncorrelated limit. As we approach the strongly 
correlated limit ($g_{12} \le 0.4$) the dependence of 
$\varepsilon_{\rm C}$ on $N_e/N_a$ is very weak even for 
$N_e/N_a \approx 1$. One concludes that $\varepsilon_{\rm C}(g_{12})$ 
is a useful  basis for introducing practical approximations 
on more complex systems.

\begin{figure}[x]
\centerline{\resizebox{8.2cm}{10.2cm}{\includegraphics{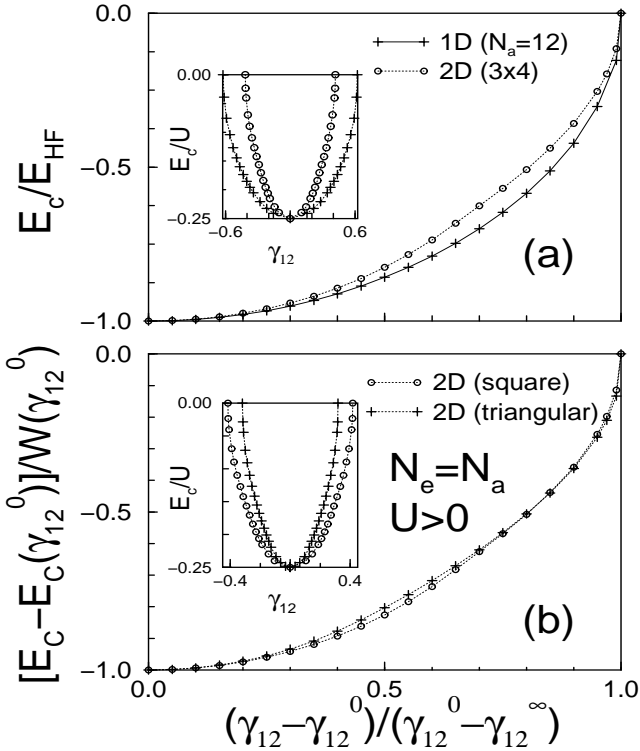}}}
\caption{
Correlation energy $E_{\rm C}(\gamma_{12})$ 
of the Hubbard model on different lattice structures.
Finite clusters with periodic boundary conditions are considered at 
half-band filling: (a) one-dimensional (1D) ring ($N_a=12$), 
2D square and triangular lattices ($3 \times 4$ clusters), and (b)
3D fcc and bcc lattices (two-tetrahedron cluster with $N_a=8$) 
\protect\cite{foot_W0}.
$U$ refers to the intra-atomic Coulomb repulsion ($U>0$). Notice the effect 
of scaling $\gamma_{12}$ with the uncorrelated $\gamma_{12}^0$ by 
comparing main and inset figures.
}
\label{fig:xcfdima}
\end{figure}

The correlation energy $E_{\rm C}$ is a universal 
functional of the complete single-particle density matrix $\gamma_{ij}$. 
$E_{\rm C}[\gamma_{ij}]$  and $W[\gamma_{ij}]$ may depend 
on $N_a$ and $N_e$ but are independent of $t_{ij}$ and in 
particular of the lattice structure. The functional 
$E_{\rm C}(\gamma_{12})$ considered in this paper depends by 
definition on the type of lattice, since the constraints imposed
in the minimization only apply to NN bonds.
In order to investigate this problem we have determined 
$E_{\rm C}(\gamma_{12})$ for 2D and 3D finite clusters having 
$N_a \le12$ sites  and periodic boundary conditions. 
In Fig.~\ref{fig:xcfdima} we compare these results 
with those of the 1D $12$-site periodic ring. As shown in the 
inset figure, the qualitative behavior is in all cases very similar. 
The main quantitative differences come from the domain of 
representability of $\gamma_{12}$, i.e., from the values of 
$\gamma_{12}^{0+}$ and $\gamma_{12}^{0-}$  
($\gamma_{12}^{0-} \le \gamma_{12 } \le \gamma_{12}^{0+}$). 
Once scaled as a function of $\gamma_{12} / \gamma_{12}^{0}$, 
$E_{\rm C}$ depends rather weakly on the lattice structure. Notice that
the Hartree-Fock energy $E_{\rm HF} = (U/4) N_a$ is the same 
for all structures. However, for the BCC structure
we obtain $W(\gamma_{12}^0) < E_{\rm HF}$, i.e., 
$E_{\rm C}(\gamma_{12}^0) <0$, due to degeneracies in the 
single-particle spectrum of the considered finite cluster
[see inset Fig.~\ref{fig:xcfdima}(b)]. In order to correct for 
this finite size effect it is here more appropriate to consider 
$\varepsilon_{\rm C} = [E_{\rm C}(\gamma_{12}) - E_{\rm C}(\gamma_{12}^0)] 
/ W(\gamma_{12}^0)$. Still, the differences in 
$\varepsilon_{\rm C}$ between BCC and FCC structures appear to 
be more important than between square and triangular 2D lattices. 
This is probably related to the degeneracies in the spectrum of the
BCC cluster, as already observed for rings with $N_e = 4m$ 
[Figs.~\ref{fig:nxcanft} and \ref{fig:nxcanfl1}(a)]. 

The largest changes in $\varepsilon_{\rm C}$ for different lattice 
structures are observed for intermediate degree of delocalization
($g_{12}\simeq 0.7$--$0.9$, see Fig.~\ref{fig:xcfdima}). 
Note that there is no monotonic trend as a function of the 
lattice dimension. For example, for $g_{12} = 0.7$--$0.9$, 
$\varepsilon_{\rm C}$ first increases somewhat as we go from 1D to 
2D lattices, but it then decreases coming close to the 1D curve for 
the 3D FCC lattice 
[$\varepsilon_{\rm C}({\rm 2D}) > 
  \varepsilon_{\rm C}({\rm FCC}) > 
  \varepsilon_{\rm C}({\rm 1D}) > 
  \varepsilon_{\rm C}({\rm BCC})$ for $0.7\le g_{12}\le 0.9$]. 
Finally, it is worth noting that in the strongly correlated limit
($g_{12}\le 0.3$) the results for $\varepsilon_{\rm C}(g_{12})$
are nearly the same for all considered lattice structures
(see Fig.~\ref{fig:xcfdima}). This should be useful in order to 
develop simple general approximations to $E_{\rm C}(\gamma_{12})$ 
in this limit.

\subsection{Attractive interaction $U<0$ }

The attractive Hubbard model describes itinerant electrons with local 
intra-atomic pairing ($U<0$). The electronic correlations are very 
different from those found in the repulsive case discussed so far. 
In particular Levy's interaction energy functional $W(\gamma_{ij})$ 
now correspond to the maximum average number of double occupation for a 
given $\gamma_{ij}$ [see Eq.~(\ref{eq:whub})]. Therefore, it is very 
interesting to investigate the properties of the correlation energy 
functional $E_{\rm C}(\gamma_{ij})$ also for $U<0$ and to contrast them with 
the results of the previous section. 

\begin{figure}[x]
\centerline{\resizebox{8.2cm}{10.2cm}{\includegraphics{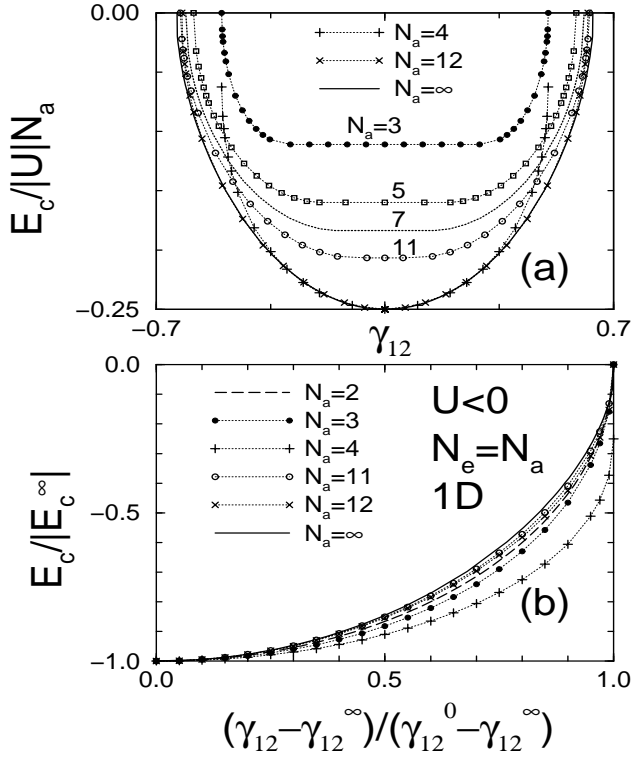}}}
\caption{ 
Correlation energy $E_{\rm C}$ of the attractive
Hubbard model ($U<0$) on one-dimensional rings with $N_a$ sites
and $N_e = N_a$ electrons.   
In (a) $E_{\rm C}$ is shown as a function of the density-matrix 
element $\gamma_{12}$ between nearest neighbors (NN), 
$\gamma_{ij} = \gamma_{12}$ for all NN $ij$.
In (b) $E_{\rm C}/|E_{\rm C}^\infty |$ is given as a function of the 
degree of delocalization 
$g_{12}=(\gamma_{12} - \gamma_{12}^\infty) / 
(\gamma_{12}^0 - \gamma_{12}^\infty)$. 
See caption of Fig.~\protect\ref{fig:nxcanfl1}.
}
\label{fig:xcsuft}
\end{figure}

\begin{figure}[x]
\centerline{\resizebox{8.2cm}{10.2cm}{\includegraphics{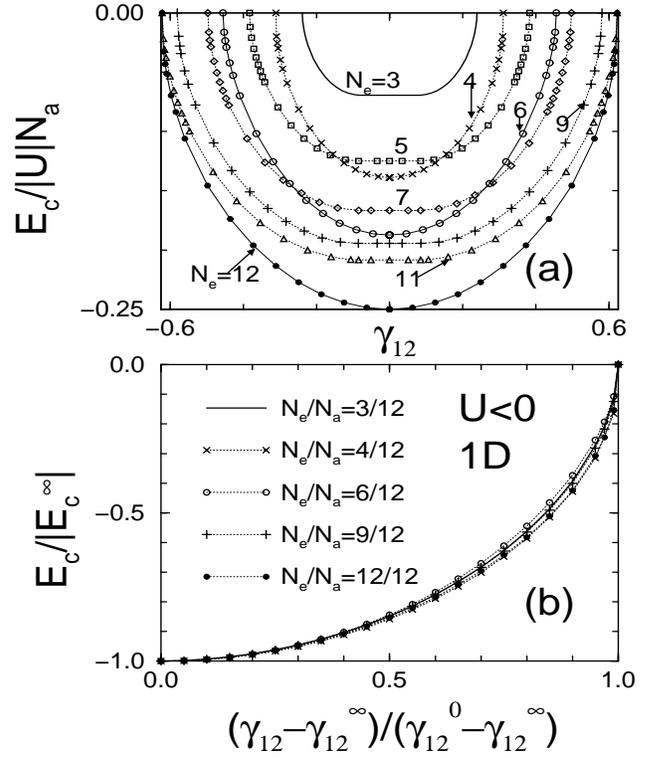}}}
\caption{ 
Band-filling dependence of the correlation energy 
$E_{\rm C}(\gamma_{12})$ of the one-dimensional attractive Hubbard model 
($U<0$). The number of sites is $N_a= 12$ and the number of electrons 
$N_e$ are indicated. 
In (a) $E_{\rm C}$ is shown as a function of $\gamma_{12}$ and 
in (b) $E_{\rm C}/|E_{\rm C}^\infty |$ is given as a function of the 
degree of delocalization 
$g_{12}=(\gamma_{12} - \gamma_{12}^\infty) / 
(\gamma_{12}^0 - \gamma_{12}^\infty)$.
$E_{\rm C}(\gamma_{12},N_e) = E_{\rm C}(-\gamma_{12}, N_e) = 
E_{\rm C}(\gamma_{12}, 2 N_a - N_e)$ \protect\cite{foot_e-hole}. 
}
\label{fig:xcsufl}
\end{figure}

In Fig.~\ref{fig:xcsuft} the correlation energy $E_{\rm C}(\gamma_{12})$ of 
the attractive Hubbard model is given at half-band filling for various 
finite rings ($N_a \le 12$) and for the infinite 1D chain ($N_e=N_a$). 
The band-filling dependence of $E_{\rm C}(\gamma_{12})$ is shown in 
Fig.~\ref{fig:xcsufl} for  a finite $12$-site ring 
($N_e \le N_a = 12$). As in the repulsive case, 
$\gamma_{12}^{0 -} \le \gamma_{12} \le \gamma_{12}^{0 +}$ since 
the domain of representability of $\gamma_{12}$ is independent of 
the form or type of the interaction. Moreover, 
$E_{\rm C}(\gamma_{12})=E_{\rm C}(-\gamma_{12})$ due to the electron-hole 
symmetry of bipartite lattices \cite{foot_e-hole}. Starting from 
$\gamma_{12}^{0+}$ or $\gamma_{12}^{0-}$ 
($\gamma_{12}^{0+}=-\gamma_{12}^{0-}=\gamma_{12}^{0}$ 
on bipartite lattices), $E_{\rm C}(\gamma_{12})$ decreases with decreasing 
$|\gamma_{12}|$ reaching the minimum $E_{\rm C}^{\infty}$ for 
$\gamma_{12}=\gamma_{12}^{\infty}$ and for 
$\gamma_{12}=\gamma_{12}^{\infty -}$ 
($\gamma_{12}^{\infty +} = -\gamma_{12}^{\infty -} = \gamma_{12}^{\infty}$ 
in this case). For $N_e$ even, $W(\gamma_{12}^{\infty})=N_eU/2$, 
and for $N_e$ odd, $W(\gamma_{12}^{\infty})=(N_e-1)U/2$, which correspond 
to the maximum number of electron pairs that can be formed. For $N_e$ even, 
the minimum $E_{\rm C}^{\infty}=U(N_e/2)[1 - N_e/(2N_a)]$  
is achieved only for a complete electron localization 
(i.e, $\gamma_{12}^{\infty}=0$). In contrast, for odd $N_e$ a finite-size 
effect is observed. In this case, one of the electrons remains unpaired even 
in the limit of strong  electron correlations and the minimum of 
$E_{\rm C}$ is $E_{\rm C}^{\infty}=U[(N_e-1)/2][1-(N_e+1)/(2N_a)]$. Moreover, 
non-vanishing $\gamma_{12}^{\infty}$ are obtained as a result of the 
delocalization  of the unpaired electron. $\gamma_{12}^{\infty}$ represents
the maximum bond order that can be obtained when $(N_e-1)/2$ electron  
pairs are formed 
($\gamma_{12}^{\infty} \rightarrow 0$ for $N_a \rightarrow \infty$,  
$N_e$ odd). Notice  that in all cases the ground state $\gamma_{12}^{gs}$ 
is found   in the interval 
$\gamma_{12}^{\infty} \le \gamma_{12}^{gs} \le \gamma_{12}^{0}$. 

It is interesting to observe that $E_{\rm C}(\gamma_{12})$ can be appropriately
scaled in a similar way as for $U>0$. In Fig.~\ref{fig:xcsufl}(b), 
$\varepsilon_{\rm C}(g_{12})=E_{\rm C}/|E_{\rm C}^{\infty}|$ is shown as a 
function of the degree of delocalization 
$g_{12}=(\gamma_{12}-\gamma_{12}^{\infty}) /
(\gamma_{12}^0-\gamma_{12}^{\infty})$. $\varepsilon_{\rm C}(g_{12})$ presents 
a  pseudo-universal behavior in the sense that it depends weakly on $N_a$ 
and $N_e$. The main deviations from the common trend are found for 
$N_e=N_a=4$. As already discussed for $U>0$, this is a consequence of  
degeneracies in the single-particle spectrum. In this case, the wave 
function corresponding to the minimum in Levy's functional for 
$\gamma_{12} \rightarrow \gamma_{12}^0$ [Eq.~(\ref{eq:whub})] cannot 
be described by a single Slater determinant and 
$W(\gamma_{12} \rightarrow \gamma_{12}^0) < E_{\rm HF}$.

\section{Conclusion}
\label{sec:conc}

Density-matrix functional theory has been applied to lattice Hamiltonians 
taking the Hubbard model as a particularly relevant example. In this framework 
the basic variable is the single-particle density matrix $\gamma_{ij}$ and 
the key unknown is the correlation energy functional $E_{\rm C}[\gamma_{ij}]$.
The challenge is therefore to determine $E_{\rm C}[\gamma_{ij}]$ or to provide 
with useful accurate approximations for it. In this paper we presented a 
systematic study of the functional dependence of $E_{\rm C}(\gamma_{12})$ on 
periodic lattices, where $\gamma_{12}$ is the density-matrix element 
between nearest neighbors ($\gamma_{ij}=\gamma_{12}$ for all NN $ij$). 
Based on finite-cluster exact diagonalizations and on the Bethe-Ansatz 
solution of the 1D chain, we derived rigorous results for 
$E_{\rm C}(\gamma_{12})$ of the  Hubbard model as a function of the number 
of sites $N_a$, band filling $N_e/N_a$ and lattice 
structure. A basis for applications  of density-matrix functional theory 
to many-body lattice models is thereby provided. The observed 
pseudo-universal behavior of 
$\varepsilon_{\rm C}(g_{12})=E_{\rm C}/E_{\rm HF}$ as a 
function of $g_{12}=(\gamma_{12}-\gamma_{12}^{\infty})/
(\gamma_{12}^{0}-\gamma_{12}^{\infty})$  encourages transferring 
$\varepsilon_{\rm C}(g_{12})$ from finite-size systems to infinite 
lattices or even to different lattice geometries. In fact, the exact 
$E_{\rm C}(\gamma_{12})$ of the Hubbard dimer has been recently 
used to infer a simple general Ansatz for 
$E_{\rm C}(\gamma_{12})$ \cite{tobe}. With this approximation to 
$E_{\rm C}(\gamma_{12})$  the ground-state energies and charge-excitation 
gaps of 1D and 2D lattices have been determined successfully in the 
whole range of $U/t$. Further investigations, for example, by considering 
magnetic impurity models or more complex multiband Hamiltonians, are
certainly worthwhile. 

\acknowledgements
The authors gratefully acknowledge the financial support provided by
CONACyT-Mexico (RLS) and by the Alexander von Humboldt Foundation (GMP).

%
%

\newpage

\end{multicols}

\begin{references}
\bibitem{foot_tij}
Given a basis set $\{\phi_{i}\}$, the hopping integrals are expressed as
%
\begin{equation}
t_{ij} = \int d^3r \; \phi_{i}^{*}(\vec r) 
[- \frac{\hbar^2}{2m} \nabla ^{2} + 
V_{ext}(\vec r)] \phi_j (\vec r) \; , 
\end{equation}
%
and the matrix elements of the Coulomb interaction as
%
\begin{equation}
V_{ijkl} = e^2 \int d^{3}r\; d^{3}r' \; 
\phi_i^*(\vec r) \phi_k^*(\vec r\;\!') 
\frac{1}{|\vec r -   \vec r\;\!'|} \phi_l(\vec r\;\!') \phi_j(\vec r)
\; .
\end{equation}
\bibitem {rep}
A single-particle density matrix $\gamma_{ij}$ is said to be pure-state 
$N$-representable if an $N$-particle state $|\Psi \rangle$ exists such that 
$\gamma_{ij} = 
\langle \Psi| \sum_\sigma \hat c_{i \sigma }^{\dagger}
\hat c_{j \sigma }|\Psi \rangle$ for all $i$ and $j$.
\bibitem {foot_ens}
An extension of the definition domain of $E[\gamma_{ij}]$ to 
ensemble-representable density matrices $\Gamma_{ij}$ is straightforward 
following the work by Valone (see Ref.~\protect\cite{val}). Ensemble density 
matrices are written as $\Gamma_{ij} = \sum_n w_n 
\langle\Psi_n| \sum_\sigma \hat c_{i \sigma }^{\dagger} \hat c_{j \sigma }
|\Psi_n \rangle$ with $w_n \ge 0$ and $\sum_n w_n = 1$. 
In practice, the $\Gamma_{ij}$ are much easier to characterize than 
pure-state density matrices.
\bibitem{foot_e-hole}
The electron-hole transformation 
$\hat h_{i\sigma}^\dagger = \hat c_{i\sigma}$ leaves the Hubbard 
Hamiltonian and Eq.~(\protect\ref{eq:evhub}) formally unchanged, 
except for an additive constant and a change of sign in the 
hopping integrals $t_{ij}$ [or of $\lambda_{ij}$ in 
Eq.~(\protect\ref{eq:evhub})] . 
$E_{\rm C}(\gamma_{12}, N_e) =  E_{\rm C}(- \gamma_{12}, 2N_a - N_e)$, 
since the form of interaction term is unaffected by 
changing electrons to holes, and since the density matrix for holes satisfies 
$\gamma^h_{ij} = 
\sum_\sigma\langle \hat h_{i\sigma}^\dagger \hat h_{j\sigma}\rangle 
= - \gamma_{ji}$. Moreover, on bipartite lattices with only NN hoppings
$E_{\rm C}(\gamma_{12}, N_e) =  E_{\rm C}(- \gamma_{12}, N_e)$, 
since the sign of the NN bond order may be changed by changing the sign of 
$\hat c_{i\sigma}$ on one of the sublattices.
\bibitem{foot_U=0}
In the presence of degeneracies in the single-particle spectrum 
of finite systems one may find that the minimum of $W[\gamma_{ij}]$ 
does not correspond to a single Slater determinant, and that 
$W(\gamma_{12} \rightarrow \gamma_{12}^{0}) < E_{\rm HF}$. 
Such a behavior is observed, for example, in rings with $N_a = N_e = 4m$. 
This is a finite-size effect which importance decreases with increasing 
$N_a$: $W(\gamma_{12}^0)/(U N_a)=0.19$, $0.23$ and $0.24$ for rings with 
$N_e=N_a=4$, $8$ and $12$, respectively. $E_{\rm HF}/(U N_a)=0.25$.
\bibitem{foot_EH}
In the nonmagnetic case the Hartree-Fock energy 
of the Hubbard model is 
$E_{\rm HF}/N_a = (U/4)(N_e/N_a)^2$ for $N_e$ even and 
$E_{\rm HF}/N_a = (U/4)(N_e/N_a)^2[1-(1/N_e)^2]$ for $N_e$ odd. 
Notice that in this model the difference between $E_{\rm HF}$ and the 
Hartree energy $E_{\rm H}/N_a=(U/2)(N_e/N_a)^2$ is only 
the self interaction.
\bibitem{foot_W0}
If $W(\gamma_{12}^0)<E_{\rm HF}$, i.e., $E_{\rm C}(\gamma_{12}^0)<0$, 
it is more appropriate to consider 
$\varepsilon_{\rm C} = 
[E_{\rm C}(\gamma_{12}) - E_{\rm C}(\gamma_{12}^0)] / W(\gamma_{12}^0)$
in order to correct for this finite size effect. In this way
$\varepsilon_{\rm C}(g_{12})$ is less sensitive to the details of
the lattice structure or cluster size.
\bibitem{foot_rep}
Using a linear combination $|\Psi\rangle$ between two 
$N$-particles states $|\Psi_a\rangle$ and  $|\Psi_b\rangle$ 
satisfying  $\langle\Psi_a|\Psi_b\rangle=0$ and  
$\langle\Psi_a| \sum_{\sigma} \hat c_{i \sigma}^{\dagger} \hat c_{j \sigma}
|\Psi_b\rangle =0$ (for instance, two states with different total 
spin $S$ or $S_z$), one may represent the NN matrix elements 
$\gamma_{12} = \gamma_{ij} = \langle\Psi| 
\sum_{\sigma} \hat c_{i \sigma}^{\dagger} \hat c_{j \sigma}|\Psi\rangle$ 
in the interval $\gamma_{12}^{a} \le \gamma_{12} \le \gamma_{12}^{b}$,  
where $\gamma_{12}^{a}$ and  $\gamma_{12}^{b}$ are the bond orders
corresponding to $|\Psi_a \rangle$ and $|\Psi_b \rangle$ 
($\gamma_{12}^{a} < \gamma_{12}^{b}$). For example, one may take 
$|\Psi_a \rangle$ as a fully localized state with maximal $S_z$ 
($\gamma_{ij}^{a}=0$, $\forall {ij}$) and $|\Psi_b\rangle$ as the state
representing $\gamma_{12}^{\infty}$ for $S_z=0$.
\end{references}
\end{document}